\documentclass[mathleft]{an}
\usepackage{graphicx}
\usepackage{times}
\begin{document}

\Pagespan{1}{}
\Yearpublication{2007}
\Yearsubmission{2007}
\Month{}
\Volume{}
\Issue{}
\DOI{}

\title{Radial velocities of five globular clusters obtained with AAOmega}

\author{P. Sz\'ekely\inst{1,2}\fnmsep\thanks{Corresponding author:
  \email{pierre@physx.u-szeged.hu}\newline}
\and  L. L. Kiss\inst{3}
\and K. Szatm\'ary\inst{1}
\and B. Cs\'ak\inst{4}
\and G. \'A. Bakos\inst{5}
\and T. R. Bedding\inst{3}
}
\titlerunning{Radial velocities of globular clusters}
\authorrunning{P. Sz\'ekely et al.}
\institute{Department of Experimental Physics \& Astronomical Observatory, University
of Szeged, H-6720 Szeged, D\'om t\'er 9., Hungary
\and
Hungarian E\"otv\"os Fellowship, School of Physics, University of Sydney, NSW 2006
Australia
\and
School of Physics, University of Sydney 2006, NSW 2006 Australia
\and Department of Optics and Quantum Electronics, University of Szeged, H-6720 Szeged, 
D\'om t\'er 9., Hungary
\and Harvard-Smithsonian Center for Astrophysics, Cambridge, MA, USA}

\received{}
\accepted{}
\publonline{later}


\abstract{Using the recently commissioned multi-object spectrograph AAOmega on the 3.9m
AAT we have obtained medium-resolution near-infrared spectra for 10,500 stars in and
around five southern globular clusters. The targets were 47~Tuc, M12, M30, M55 and NGC
288. We have measured radial velocities to $\pm$1 km~s$^{-1}$ with the cross
correlation method and estimated metallicity, effective temperature, surface gravity and
rotational velocity for each star by fitting synthetic model spectra. An analysis of the
velocity maps and velocity dispersion of member stars revealed systemic rotation in four
of the target clusters.}

\maketitle

\section{Introduction} Globular clusters are among the oldest objects in our Galaxy and
their stars provide unique information on Galactic evolution through cosmic times. These
clusters are sensitive indicators of the Galactic gravitational potential because the
velocity distribution of the stars in their outskirts is affected by the ambient field of the Milky Way.
They are believed to have undergone substantial dynamical evolution, which is affected by
processes responsible for the ``evaporation'' of stars (Meylan \& Heggie 1997): tidal
interaction with the Galaxy and two-body relaxation. In most of the cases the cluster moves around the Galactic center along very elongated orbit. Close to the perigalactic position the cluster suffers the strongest gravitational interactions which can result in tidal shocks and lead to the formation of tidal tails.
Recent investigations have found long
tidal tails in two low-concentration clusters, Palomar 5 and NGC 5466, thus confirming
theoretical predictions (Odenkirchen et al. 2001;\\
Grillmair \& Johnson 2006). The
success behind these results was secured by deep photometric surveys, most notably the
SDSS project, which allowed cluster membership determination with statistical analysis of
colors and magnitudes of stars. Another possibility is offered by spectroscopic
observations of individual stars in order to derive  metallicity and radial velocity,
both efficient marker of cluster members.

AAOmega, the newly commissioned multi-object spectrograph on the 3.9 meter
Anglo-Australian Telescope provides an excellent opportunity to measure up to 350-360
radial velocities with a single exposure. Due to its large field of view (2 degrees) it
can observe stars which are located far from the cluster in the plane of the sky. With a
velocimetric accuracy of about 1-2 km~s$^{-1}$, one can easily identify escaped stars
which have the same or very similar radial velocities as the host cluster.

In this paper we present an analysis of radial velocities of cluster member stars in
terms of systemic rotation for five southern globular clusters. A detailed investigation
of their possible tidal tails has been reported by Kiss et al. (2007).

\section{Observations and data reduction}

The targets were selected from the globular cluster catalogue of Harris (1996) based on
the following criteria: we preferred nearby, unreddened clusters with large radial
velocities in respect to the Galactic field containing mostly 
disk stars. The latter is very useful to distinguish
cluster member stars. We also considered recent results on tidal tails (in case of NGC
288; Leon et al. 2000), interesting dynamical history (M12; de Marchi et al. 2006) or
internal structures (47~Tuc; Meylan \& Mayor 1986) to choose targets.  We ended up with
the following clusters: 47~Tuc, NGC 288, M12, M30, and M55. By chance we also recorded
radial velocities of a few stars in the extragalactic globular cluster NGC 121 that
belongs to the Small Magellanic Cloud. 

\begin{table}
\caption{Total number of stars identified as members, cluster radial velocity 
from Harris (1996) and the measured mean velocity.}
\label{data}
\centering
\setlength{\tabcolsep}{2mm}
\begin{tabular} {lcrr}
\hline
\hline
\noalign{\smallskip}
Cluster & No. & $v_{\rm rad}$ (H96) & $\langle v_{\rm rad}\rangle$\\
 &  stars & km~s$^{-1}$ & km~s$^{-1}$\\
\hline
\noalign{\smallskip}
47~Tuc (NGC 104) & 911 & $-$18.7 & $-$16.1\\
M12 (NGC 6218) & 158 & $-$42.2 & $-$40.4 \\
M30 (NGC 7099) & 129 & $-$181.9 & $-$178.1 \\
M55 (NGC 6809) & 433 & 174.8 & 171.7  \\
NGC 288 & 123 & $-$46.6 & $-$43.5  \\
\hline
\end{tabular}
\end{table}

Our observations were carried out on 7 nights in August, 2006. Throughout the run the
seeing stayed between 1.2 and 2.2 arcsec. In total we took spectra for more than 10,500
stars using the D1700 grating, recording near-infrared spectra centered on the Ca II
triplet lines. The spectra ranged from 8350 \AA\ to 8790 \AA\ with a resolving power
$\lambda/\Delta\lambda=10,500$. For each cluster we observed several fields centered on
the cluster, containing 300-350 stars per configuration. To reach a signal-to-noise ratio
between 50 and 250 we exposed 60 to 90 minutes in total. For sky background measurements
and guiding we used 30-40 fibers per configuration.

The target stars were selected from the 2MASS point source catalogue (Skrutskie et al.,
2006). We fitted a line to the Red Giant Branch (RGB) of each cluster and selected 
stars which matched its color and magnitude. For M12 and 47~Tuc we selected the lower
part of the RGB. Due to the sparse field star population around  M30 and NGC 288, we did
not filtered their stars, but used the whole field. The full magnitude range of the
target stars in $K$-band was 7 mag (from 8 mag to 15 mag) but for a single configuration
field we limited the brightness range to 3 mag in order to avoid cross-talk between the
fibers due to scattered light.

The spectra were reduced with the standard 2dF data processing pipeline  ({\it
drcontrol}), which extracts automatica\-lly the wavelength calibrated spectra. Continuum
normalization was done with the IRAF\footnote{IRAF is distributed by the National Optical
Astronomy Observatories, which are operated by the Association of Universities for
Research in Astronomy, Inc., under cooperative agreement with the National Science
Foundation.} task {\it onedspec.continuum}. After this we cleaned the spectra of the
remnants of the strongest skylines via linear interpolation of the neighboring continuum.

For this study we derived two parameters for each star based on its spectrum: radial
velocity and the full equivalent width of the Ca II triplet lines
($\Sigma EW=W_{8498}+W_{8542}+W_{8662}$). The latter was determined by fitting a sum
of a Lorentzian and a Gaussian to the line profiles (Cole et al. 2004). Radial
velocities were determined in an iterative way. An initial velocity estimate was provided
by the line profile fit, which was used to find the best-fit spectrum in the extensive
spectrum library of Munari et al. (2005). The fitted model spectrum was then
cross-correlated with the observed one, giving the finally adopted velocity. The
estimated accuracy is about $\pm$1-2 km~s$^{-1}$.

\section{Results}

We identified cluster members using the radial velocities, equivalent widths and $K$
magnitudes. The selection was ba\-sed on the $v_{rad}$ vs. $\Sigma EW$ and $\Sigma EW$ vs.
$K$ magnitude correlations and led to the identifications of 123-911 stars as members in
the five clusters (see Table~\ref{data}). Note that these sets are more constrained
than those of Kiss et al. (2007), who only used the radial velocity to determine cluster
membership.

\begin{figure}
\includegraphics[width=8cm]{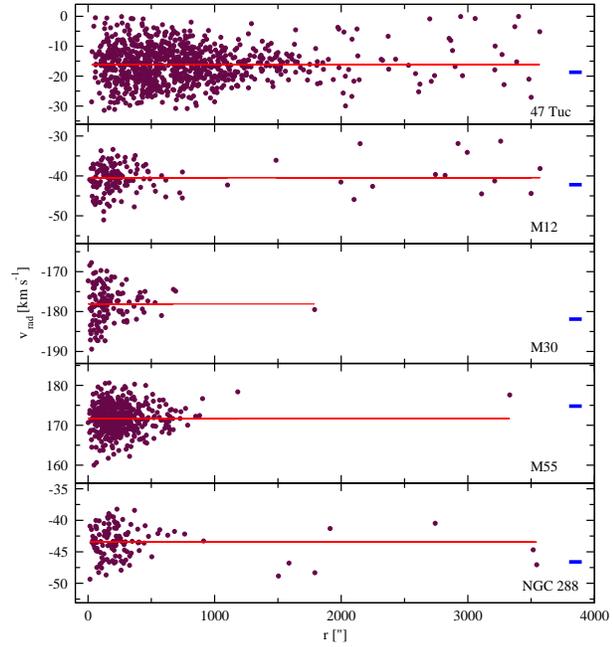}
\caption{Heliocentric radial velocities of the member stars versus distance from the 
cluster center. The horizontal lines show the mean cluster velocities. Thick lines
indicate the catalogued values from Harris (1996). The wide distribution of the velocities around the mean values is partly due to the intrinsic velocity distribution of cluster stars.}
\label{vr}
\end{figure}

First we plot heliocentric radial velocities of the member stars as function of distance
from the cluster center in Fig.~\ref{vr}. Here the horizontal lines represent the cluster
mean radial velocities, while on the right-hand side of the plot thick lines mark the
catalogued values from Harris (1996). While there are differences up to 3 km~s$^{-1}$ in
the measured means and the catalogued values, our samples are more extensive than any
previous ones for all clusters except 47~Tuc, which may explain most of the
differences. Because of the excellent statistics, 
Fig.~\ref{vr} is clearly dominated by the characteristic wedge-shaped velocity
distribution that can be used to derive the velocity dispersion profile, which is a
sensitive indicator of the dynamical properties.

\begin{figure*}
\includegraphics[width=8cm]{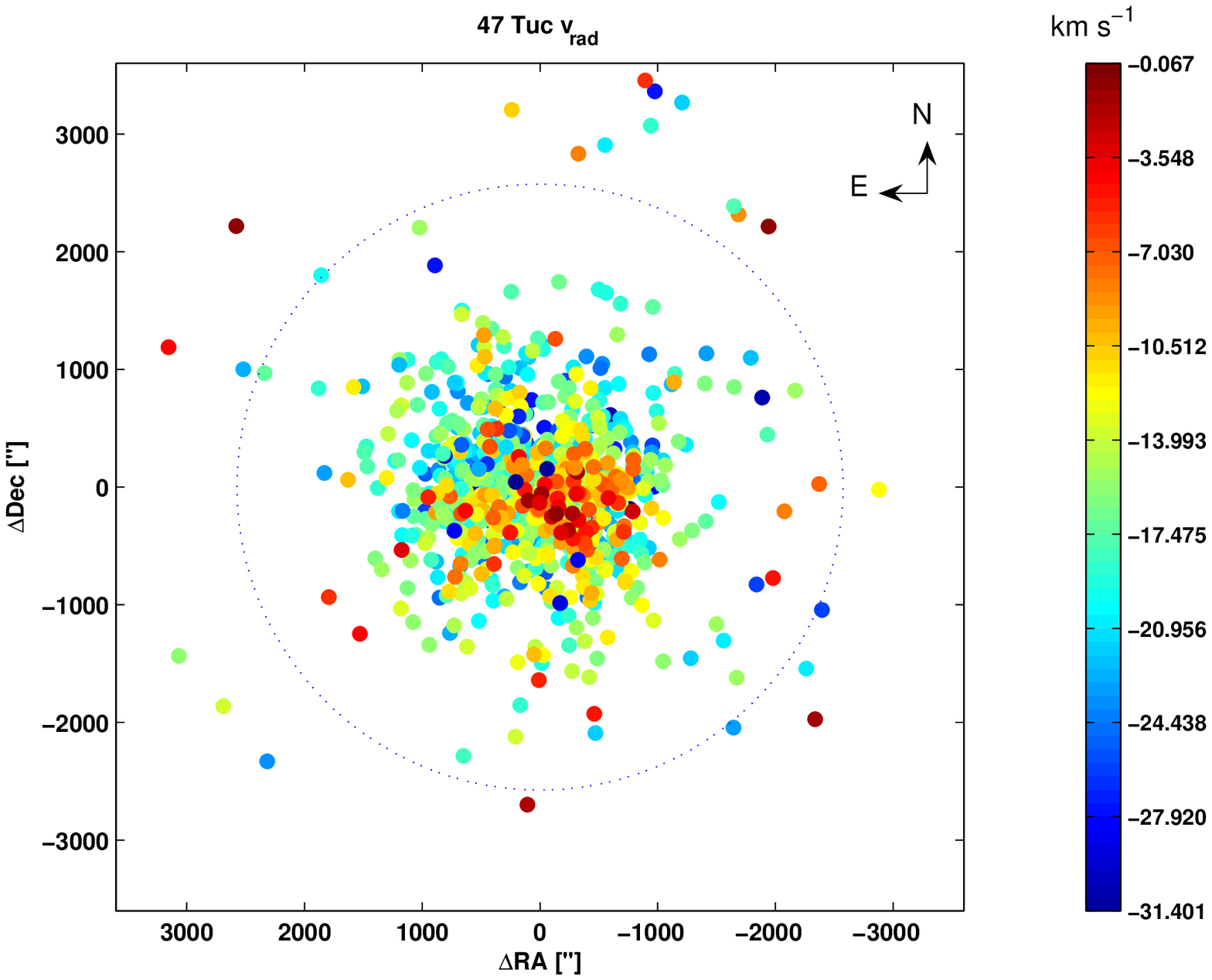}
\hskip2mm
\includegraphics[width=8cm]{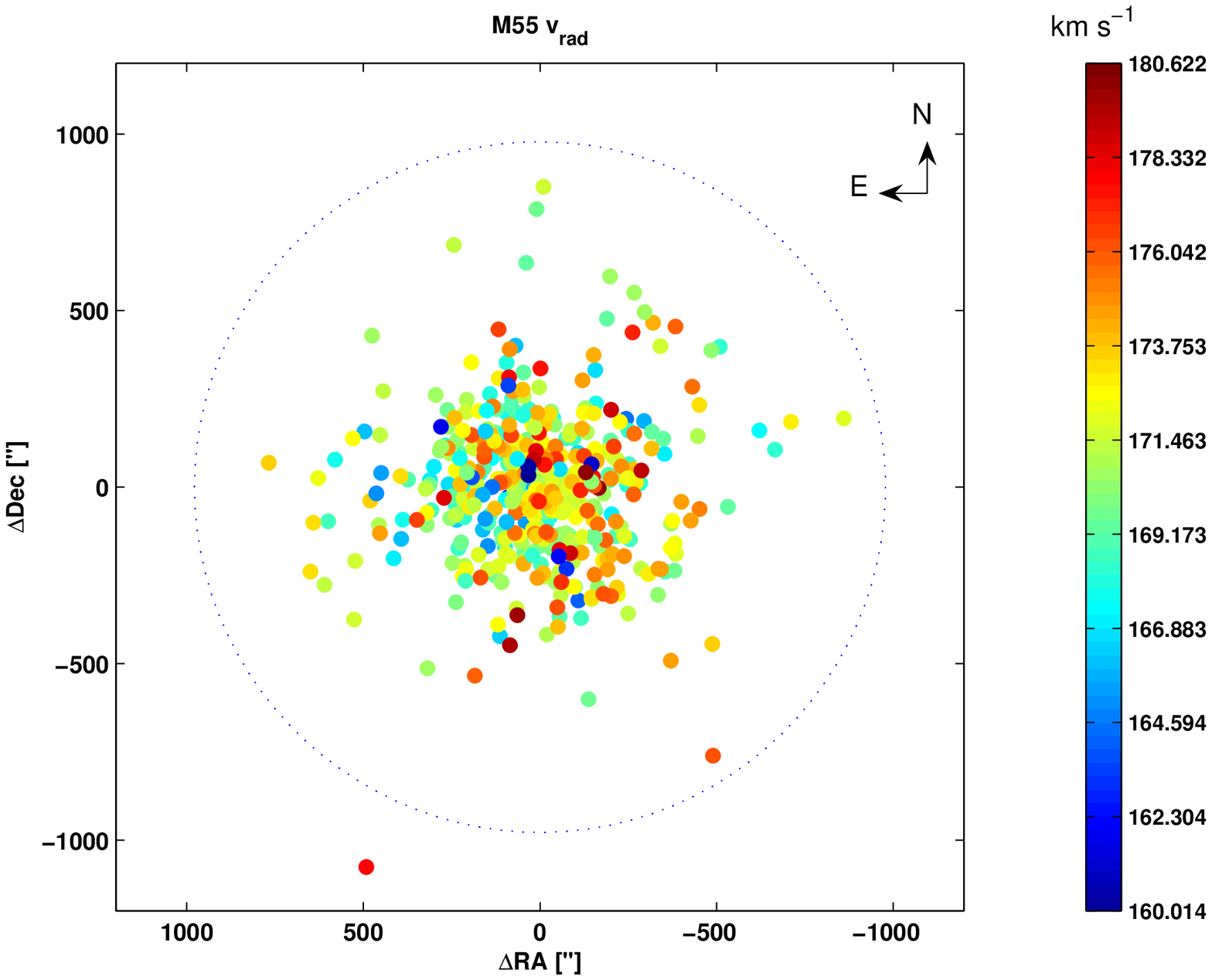}
\caption{{\it Left panel:} stellar positions in 47~Tuc with radial velocities 
encoded in colors. The circle shows the tidal radius ($r_t=42.9^\prime$). {\it Right
panel:} the same for M55 ($r_t=16.3^\prime$). Note the smaller field of view.}
\label{47tucvr}
\end{figure*}

In Fig.~\ref{47tucvr} we plot celestial positions of member stars for 47~Tuc and M55,
color-coded by their radial velocities. It is very prominent that there is a well-defined
asymmetry in both clusters, which can be interpreted as result of rotation.

\begin{figure*}
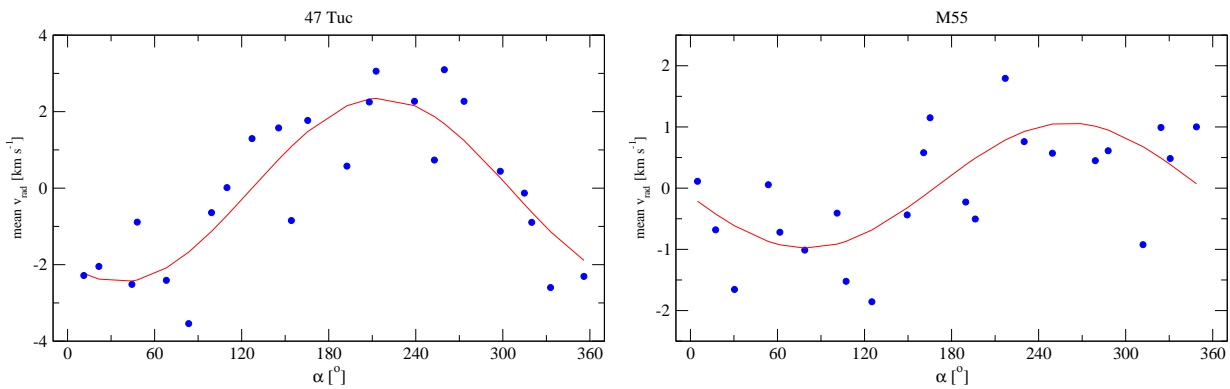

\includegraphics[width=8cm]{szekely_fig4.eps}
\hskip2mm
\includegraphics[width=8cm]{szekely_fig5.eps}
\caption{Mean radial velocities versus polar angle in 24 azimuthal bins with 
equal width. We also plotted the best-fit sine curves.}
\label{47tucmean}
\end{figure*}

To find quantitative parameters of the systemic rotation in these two clusters,  we
plotted the mean radial velocities versus position angle in Fig.~\ref{47tucmean}. The
effects of  the internal velocity dispersion were minimized by binning the velocities in 
15 degree-wide sectors of the position angles ($\alpha$=0$^\circ$ corresponding to North,
90$^\circ$ to East). The mean cluster velocities listed in Table~\ref{data} were
subtracted for clarity. The sinusoidal shape of the distributions is well recognizable in
both cases, though it is much clearer for 47~Tuc. For the other three clusters similar
plots did not reveal obvious patterns.

\begin{table}
\caption{Parameters of the detected systemic rotations: A is the peak-to-peak velocity
amplitude, while the position angle of the rotational axis is given in both equatorial 
and galactic coordinates.}
\label{data2}
\centering
\begin{tabular} {cccc}
\hline
\hline
\noalign{\smallskip}
Cluster & A & PA (eq.) & PA (gal.) \\
& km~s$^{-1}$ & $^\circ$ & $^\circ$ \\
\hline
\noalign{\smallskip}
47~Tuc & 6.6$\pm0.07$ & 130/310$\pm5$ & 122/302$\pm5$\\
M12  & -- & -- & -- \\
M30 & 1.5$\pm0.3$ & 16/196$\pm15$ & 90/270$\pm15$ \\
M55  & 2.4$\pm0.1$ & 170/350$\pm7$ & 61/241$\pm7$ \\
NGC 288 & 1.4$\pm0.15$ & 97/277$\pm10$ & 52/232$\pm10$ \\
\hline
\end{tabular}
\end{table}

\begin{figure}
\includegraphics[width=8cm]{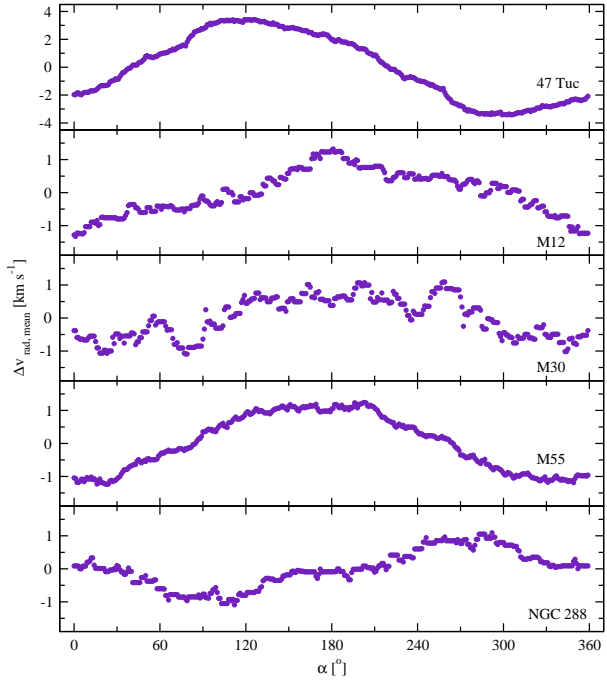}
\caption{Mean radial velocities on both sides of an imaginary axis stepped 
through the clusters. Note the different scale on the $y$ axis in case of 47~Tuc.}
\label{axis3}
\end{figure}

Another useful method to detect azimuthal dependence of the radial velocities is stepping
an imaginary axis through the globular cluster in small angular increments (e.g. one
degree) and calculate the differences of the mean or median radial velocities on both
sides of the axis (e.g. C\^ot\'e et al. 1995). The resulting curves are plotted in
Fig.~\ref{axis3}. Based on this we can put an upper limit to the rotational velocity projected into the line of sight for each cluster by considering the amplitude of the curve. For 47~Tuc, the rotational velocity is 6.6 km~s$^{-1}$, being in perfect agreement with the results of Meylan \&
Mayor (1986) and Anderson \& King (2003). The rotational velocity estimates for the other
three clusters are listed in the second column of Table~\ref{data2}. For M12, the curve is
clearly non-sinusoidal, so that we did not attempt to fit the data.

To derive the orientations of the projected rotational axis in the plane of the sky we
fitted a sine wave to the curves presented in Fig.~\ref{axis3}. The results are
listed in Table~\ref{data2}. We also converted the equatorial position angles to galactic
coordinates to see if there is a pattern in the directions. Indeed, for the four clear
detections of systemic rotation we find directions being close (within $\pm$20-25
deg) to parallel with the Galactic plane, which {\it might} be an evidence of non-random
distribution of rotational axes, but the size of this sample is too small even for some statictical approaches. In any case, we confirmed rotation in 47~Tuc and 
discovered it in M30, M55 and NGC 288. For M12 we will have to clean further the sample 
of the Galactic field contamination before a new attempt at detecting rotation is 
considered.

\section{Future work}

Our data represent a unique source of information on the studied five globular clusters.
We will separate cluster members from the Galactic field stars by detecting clumpiness in
the full parameter space of the physical parameters, including radial velocity, metallicity,
effective temperature, surface gravity. Global kinematics, mass-to-light ratios and
star formation histories will be constrained from the cleaned samples of member stars. We
will be particularly interested in the age-metallicity relation because a spread in
metallicity can be a sign of prolonged star formation over 2-4 Gyrs  (Stanford et al.
2006).

We are specifically interested in investigating mechanisms that affect velocity 
distributions in globular clusters and,
in particular, the tidal tails. Theories to be tested include tidal heating of the
evaporated stars by the external gravitational field (Drukier et al. 1998), the 
presence of a dark matter halo around the clusters (Carraro \& Lia 2000), and  
a breakdown of the Newtonian dynamics in the weak-acceleration
regime (Scarpa et al. 2007). The latter hypothesis is particularly
interesting because modified Newtonian dynamics, valid for accelerations below 
$a_0\sim 1.2\times10^{-8}$ cm~s$^{-2}$, may offer an alternative to the dark matter,
with far-reaching implications for cosmology.

For each cluster we aim to measure the  velocity  dispersion profile of stars, which
shows  a characteristic decrease and flattening at distances from the cluster centre 
(Scarpa et al. 2007). Globular clusters are relatively simple  stellar systems, with each
star's motion dictated by the Newtonian gravitational pull of the other cluster members.
However, any departure  from the classical laws of dynamics (such as Modified Newtonian 
Dynamics: MOND, originally introduced by Milgrom 1983) predicts that the flattening
should  occur at  the same  absolute acceleration regardless of the Galactic environment
of the cluster. Hence, we will characterize  the dynamical  properties of the clusters 
as a fundamental probe of Newtonian gravity.  Since rotation of the cluster can
introduce systematic errors in  the interpretation, we will subtract a smoothed velocity
field from the data. The residuals will show  whether the  velocity dispersion  decreases
at  large  radii and whether  it reverses  to an  increase  outside the  tidal radius. To
disentangle possible  breakdown  of the  Newtonian dynamics  and conventional dynamic 
effects such as  tidal heating, one has to observe clusters of very different parameters, including  different Galactocentric  distances  and determine  the  actual
acceleration at which the dispersion profile flattens. This presented sample is a good starting point in this direction and while much effort will be devoted
to model the cluster dynamics, we also plan to extend the observational data base  
with further clusters.

\acknowledgement

This project has been supported by the Hungarian OTKA Grant \#T042509, a Hungarian
E\"otv\"os Fellowship to PSz and the Australian Research  Council. LLK is supported by a
University of Sydney Postdoctoral Research  Fellowship. Support for program number
HST-HF-01170.01-A to G.\'A.B. was provided by NASA through a Hubble Fellowship grant from
the Space Telescope Science Institute, which is operated by the AURA, Inc., under NASA
contract  NAS5\-26555. G.\'A.B. also wishes to thank useful discussions to A.~P\'al. We
are very grateful to the staff of the Anglo-Australian Observatory for their kind and
helpful support during our observing run. 

\end{document}